\documentstyle[prb,aps]{revtex}

\input epsfig.sty

\begin{document}
\title{{\tt to appear in Phys. Rev. B } \\
Current-voltage characteristics of diluted Josephson-junction arrays:
scaling behavior at current and percolation threshold.}
\author{Enzo Granato}
\address{Laborat\'orio Associado de Sensores e Materiais, \\
Instituto Nacional de Pesquisas Espaciais, \\
12201 - S\~ao Jos\'e dos Campos, SP, Brazil }
\author{Daniel Dom\'{\i}nguez}
\address{Centro At\'{o}mico Bariloche, \\
8400 S. C. de Bariloche, Rio Negro, Argentina}
\maketitle
\draft

\begin{abstract}
Dynamical simulations and scaling arguments are used to study the
current-voltage (IV) characteristics of a two-dimensional model of
resistively shunted Josephson-junction arrays in presence of percolative
disorder, at zero external field. Two different limits of the
Josephson-coupling concentration $p$ are considered, where $p_c$ is the
percolation threshold. For $p$ $>$ $p_c$ and zero temperature, the IV curves
show power-law behavior above a disorder dependent critical current. The
power-law behavior and critical exponents are consistent with a simple
scaling analysis. At $p_c$ and finite temperature $T$, the results show the
scaling behavior of a $T=0$ superconducting transition. The resistance is
linear but vanishes for decreasing $T$ with an apparent exponential
behavior. Crossover to non-linearity appears at currents proportional to $%
T^{1+\nu _T}$, with a thermal-correlation length exponent $\nu _T$
consistent with the corresponding value for the diluted XY model at $p_c$.
\end{abstract}


\section{Introduction}

There has recently been an interest, both experimental \cite
{garland,martinoli} and theoretical \cite
{lubensky,bradley,leath0,leath,stroud1,stroud2,teitel,stroud_mc,kiss}, in
the resistive behavior of diluted Josephson-junction arrays (JJA). These
systems provide a useful model for several universal transport properties of
granular high-$T_c$ materials \cite{prester} and can also be physically
realized as superconducting arrays \cite{garland,martinoli} or wire networks 
\cite{networks} with accurately controlled parameters using microfabrication
techniques. Although most investigations have considered the combined
effects of disorder and magnetic fields which leads to the vortex glass
state \cite{fisher}, there are also interesting questions even in the
absence of an external field as a result of disorder. The effects of
percolative disorder on the resistive transition and nonlinear
current-voltage (IV) relation, at finite temperatures, have been studied in
artificial two-dimensional arrays \cite{garland} and also in the context of
the high-$T_c$ oxides \cite{onogi}. For an ordered two-dimensional JJA,
which is isomorphic to the XY model, it is well-known that the resistive
transition is in the Kosterlitz-Thouless (KT) universality class where the
low-temperature phase has a nonlinear IV relation, $V\propto I^{\tilde a(T)}$, due
to current induced vortex-pair unbinding. The temperature dependent exponent 
$\tilde a(T)$ decreases with increasing temperature and reaches $\tilde a(T_c)=3$ at the
transition \cite{lobb}. In presence of percolative disorder but well above
the percolation threshold, this behavior still persists with a broaden
resistive transition and disorder renormalized critical temperature $T_c(p)$
and exponent $\tilde a(T,p)$, where $p$ is the fraction of superconducting grains 
\cite{garland}. Right at the percolation threshold $p_c$, however, as for
the XY model \cite{stinchcombe}, the transition temperature is expected to
vanish but a divergent correlation length $\xi \sim T^{-\nu _T}$ and thermal
critical exponent $\nu _T$ can still be defined for $T>0$. A natural
question arises as to what should be  the behavior of the IV curves for $T>0$ in
this limit since the correlation length is finite. In fact, one expects the
increasing correlation length as $T\rightarrow 0$ to have important
consequences at finite $T$ for the nonlinear behavior of the IV curves \cite
{fisher,eg}. The linear resistance is expected to be finite for any $T>0$
but crossover to non-linearity can appear for increasing currents due to the
finite correlation length with a universal behavior determined by the $T=0$
transition. It is important to understand this behavior in detail because,
in two dimensions and in presence of an external magnetic field, an
additional $T=0$ (vortex-glass like) transition is expected \cite{lubensky}
near $p_c$ which should also have a similar effect but it could be difficult
to separate these two effects experimentally. Also, in principle, the
universal behavior at $p_c$ may be useful to identify the proximity to the
percolation threshold in systems where $p_c$ is uncertain as in granular
materials modeled as diluted JJA.

There are also other interesting effects of disorder on the current-voltage
relation which appear in a different limit and are not well understood.
Above the percolation threshold $p_c$ and zero temperature, the
IV characteristics shows a power-law behavior, $V\sim
(J-J_c)^a$, above a critical current density $J_c$. For example, an exponent 
$a\sim 3 $ was obtained  from numerical simulations by Xia and Leath \cite{leath} 
in $d=2$. Clearly, this behavior at $T=0$ and finite current density $J_c$ is 
unrelated to the
condition $\tilde a(T_c)=3$ at the KT transition mentioned above and requires
different considerations. In a linearized model of the critical current as a
phase-coherence breakdown phenomenon \cite{leath}, it has been argued that the
critical current of a diluted array vanishes in the thermodynamic limit. In
this case, the apparent finite $J_c$ and the exponent $a$, presumably
universal, are in fact an artifact of finite system sizes. However, it is
possible that nonlinearities in the array problem could lead to a finite
value even in the large system limit \cite{stroud1,stroud2}. In the latter
scenario $J_c$ is finite but is expected to vanish at the percolation
threshold as $J_c\sim (p-p_c)^v$ with an exponent $v=\nu _p(d-1)$, where $%
\nu _p$ is the percolation exponent and $d$ is the dimensionality of the
system. Unlike the behavior at finite temperatures, little is known however
about the precise value of $a$ and the universality class of the
current-induced transition leading to the power-law behavior for $p>p_c$,
under the assumption that there is in fact a well defined $J_c$.
Nevertheless, it is important to understand these features in detail in
order to be able to separate these effects, which arise only from disorder,
from others induced by a combination of disorder and strong magnetic  fields 
as in the case of the vortex glass which also leads to values of $a$ of the same
order \cite{dd} at $T=0$. Indeed, in experimental systems, these effects can
appear simultaneously and it is possible that some of the power-law behavior
observed in the IV characteristics of granular high-$T_c$ materials that
tend to be regarded as a manifestation of vortex glass behavior is actually
a result of intrinsic geometrical disorder and would persist even in the
absence of the external field.

In this work, we present dynamical simulations of the IV characteristics of
diluted two-dimensional JJA, at zero external field, in the two different
limits discussed above which allow for a scaling analysis in terms of a
single length scale. (i) At $T=0$ and $p>p_c$, we study the power-law
behavior in the IV curves above $J_c$ as resulting from a current driven
dynamical transition. We discuss scaling arguments for this behavior which
leads to $a=\nu _I(z+1)$, where $\nu _I$ is a correlation length exponent, $%
z $ a dynamical exponent and compare with numerical results. (ii) At the
percolation threshold $p=p_c$ and finite $T$ we study the scaling of the IV
curves as resulting from a $T=0$ superconducting transition. The results are
consistent with a linear resistance at finite temperatures and nonlinear
behavior appearing at current densities proportional to $T^{1+\nu _T}$,
where $\nu _T$ is the thermal-correlation length exponent for the diluted XY
model at $p_c$.

\section{Model and Simulation}

We consider a model of resistively shunted JJA consisting of coupled
superconducting islands located at the nodes of a square network with
Josephson and normal currents flowing between them. The nodes are located at 
${\bf r}=m\hat{x}+n\hat{y}$ with unit lattice constant. The current $I_{{\mu 
}}({\bf r})$ flowing between ${\bf r}$ and ${\bf r}+{\hat{\mu}}$, is modeled
as \cite{leath,dd,algor} 
\begin{equation}
I_\mu ({\bf r})=I_{{\bf r}\mu }^0\sin (\Delta _\mu \theta ({\bf r},t))\,+\,%
\frac \hbar {2eR_{{\bf r}\mu }}\frac{d\Delta _\mu \theta ({\bf r},t)}{dt}%
\,+\,{\eta }_\mu ({\bf r},t).  \label{rsj}
\end{equation}
Here $\Delta _\mu \theta ({\bf r},t)$ is a discrete gradient of the
superconducting phases $\theta ({\bf r},t)$, $I_{{\bf r}\mu }^0$ is the
critical current of the junctions, and $R_{{\bf r}\mu }$ is a shunt
resistance between the islands. The white noise random variable $\eta _\mu (%
{\bf r},t)$ represents thermal Johnson fluctuations in the current with
covariance, 
\begin{equation}
\langle \eta _\mu ({\bf r},t)\eta _{\mu ^{\prime }}({\bf r^{\prime }}%
,t^{\prime })\rangle =\frac{2kT}{R_{{\bf r}\mu }}\delta _{{\bf r},{\bf %
r^{\prime }}}\delta _{\mu ,\mu^\prime}\delta (t-t^{\prime }).  \label{noise}
\end{equation}
We assume that disorder affects only the Josephson coupling. In the absence
of the coupling, $I_{{\bf r}\mu }^0=0$, and so only normal current flow
between the neighboring grains. Disorder effects in $R_{{\bf r}\mu }$ are
assumed to be less important. This model has been applied to the study of
transport properties of high-$T_c$ oxides \cite{onogi} and, in its site
dilution version, to composite superconductors \cite{stroud2}. For proximity
coupled artificial arrays \cite{garland,martinoli}, this model is also a
reasonable approximation since, as pointed out in Ref. \onlinecite{martinoli}%
, the underlying normal-conducting layer over which the superconducting
grains are deposited provides a roughly constant $R_\mu $. Dilution of the
junctions is introduced by taking critical currents $I_{{\bf r}\mu }^0=I_0$
with probability $p$ or $I_{{\bf r}\mu }^0=0$ with probability $1-p$ and
constant shunt resistance, $R_\mu =R_0$. After combining Eqs.~(\ref{rsj})
with current conservation at each node, $\Delta _\mu \cdot I_\mu ({\bf r}%
)=I^{ext}({\bf r})$, the dynamical equation of motion for $\theta (r,t)$
becomes, 
\begin{equation}
\dot{\theta}({\bf r},t)=-\sum_{{\bf r^{\prime }}}G({\bf r},{\bf r^{\prime }}%
)\left\{ I^{ext}({\bf r^{\prime }})-\Delta _\mu \cdot \left[ I_{{\bf %
r^{\prime }}\mu }^0\sin (\Delta _\mu \theta ({\bf r^{\prime }},t))+\eta _\mu
({\bf r^{\prime }},t)\right] \right\} ,  \label{green}
\end{equation}
with $G({\bf r},{\bf r^{\prime }})$ the $d=2$ lattice Green function.
Dimensionless quantities are used with time in units of $\tau _J=\hbar
/2eR_0I_0$, currents in units of $I_0$, voltages in units of $R_0I_0$ and
temperatures in units of $\hbar I_0/2ek_B$. We choose periodic boundary
conditions along the $x$-direction and open boundary conditions along the $y$%
-direction. The array has $L\times L$ bonds, corresponding to $L\times (L+1)$
nodes. The total current $I$ is injected uniformly along the y-direction
with $i^{ext}(m,n)=J(\delta _{n,1}-\delta _{n,L+1})$ where $J$ is the
current density $J=I/L$ . Eqs.~(\ref{green}) are solved with the second
order Runge-Kutta-Helfand-Greenside algorithm for stochastic differential
equations \cite{algor} with a time step of $\Delta t=0.1\tau _J$. Temporal
averages are taken over a time of $2000\tau _J$ after a transient time of $%
500\tau _J$. The matrix multiplication by $G({\bf r},{\bf r^{\prime }})$ is
performed by means of a fast Fourier transform and cyclic reduction
algorithm as used in Ref.~\onlinecite{dd,algor}. The voltage drop along the $%
y$-direction is given by 
\begin{equation}
V=\frac 1L\sum_{m=1}^L\langle \dot{\theta}(m,n=L+1)-\dot{\theta}%
(m,n=1)\rangle 
\end{equation}
in dimensionless units, where $\langle \ldots \rangle $ is a time average,
and the average electric field is given by $E=V/L$ . 

\section{Results and Discussion}

Figure 1 shows some of the IV characteristics obtained numerically at $T=0$.
Our averaging time is a factor of two larger than the one used in Ref. %
\onlinecite{stroud2} to study a site dilution version of the same model in $%
d=3$. As a test of the numerical method we included in Figure 1 the
calculation for $p=1$ where the array behaves as a single junction with a
critical current density $J_c= 1$. Above the percolation threshold $p_c=1/2$%
, there is an infinite cluster of superconducting junctions through the
system and an apparent finite critical current density $J_c$ below which the
voltage is very small. For $p < p_c$, only isolated finite clusters occurs
and the resistance is Ohmic for small currents.

\begin{figure}[h]
\centering\epsfig{file=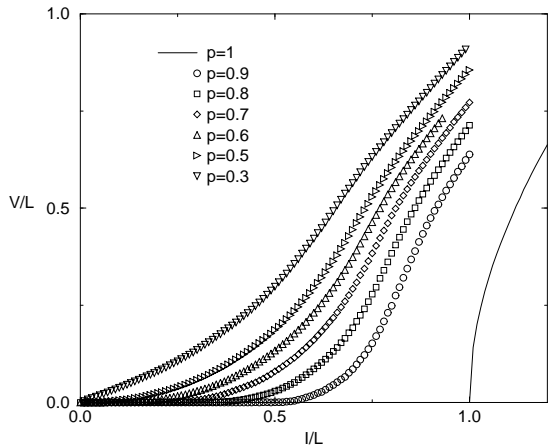,bbllx=1cm,bblly=1cm,bburx=20cm,
bbury=20cm,width=8cm}

\caption{Current-voltage characteristics for a $L= 64$ array, at zero
temperature, as a function of the Josephson-coupling concentration $p$}
\end{figure}

Figure 2a shows the behavior of the critical current density $J_c$ and
linear resistance $R_L = \lim_{J\rightarrow 0} E/J$ as a function of $p$ at
zero temperature, for the largest system size used in the simulations, $L=64$%
. $J_c$ decreases with $p$ and presumably vanishes at the percolation
threshold $p_c$ while $R_L$ is nonzero only for $p < p_c$ and also appears
to vanish at $p_c$. Unfortunately, closer to $p_c$ our data is not accurate
enough to test the expected power law behavior \cite{stroud1,stroud2} for
the critical current $J_c(p)=c\ (p-p_c)^v $, with $v=\nu_p(d-1)=4/3$, and
similar behavior for the linear resistance $R_L \sim (p_c-p)^s$, with $s
\sim 1.3$.

The exponent $a$ obtained from a power-law fit to the current-voltage
characteristics just above $J_c$, $E\sim (J-J_c)^a$, as a function of $p$ is
indicated in Fig. 2b. At $p=1$ the system behaves as a single junction and $%
a=1/2$ exactly \cite{bishop} which agrees with the numerical simulation. In
presence of disorder for decreasing $p$, this exponent jumps to a roughly
constant value, $a\sim 2.5(2)$, for the largest system size $L=64$. For the
smaller system size its $p$-dependence is more significant but we consider
this as an effect of small system sizes where the true asymptotic limit has
not been reached. The value of $a$ for $L=32$ is in fair agreement with
other simulations \cite{leath} at a fixed value of $p=0.9$ for comparable
system sizes even though the model used differs from ours in that dilution
affects both the Josephson coupling and the associated shunt resistance
simultaneously. We believe that the difference between the models should not
affect the behavior of $a$ far above $p_c$ where there is an infinite
cluster of superconducting junctions and only finite clusters of
nonsuperconducting junctions. A more accurate estimate of $ a $ would require 
a precise  determination of $J_c$, many averages over disorder and long
simulations due to  the divergent relaxation time near $J_c$ as discussed below.
Despite the uncertainties in the estimate of $a$, the behavior in Fig. 2b for 
the largest system size suggests that $a$ could be a universal critical exponent
independent of the degree of disorder parametrized by $p$ as long as $%
p_c<p<1 $.

\begin{figure}[tbp]
\centering\epsfig{file=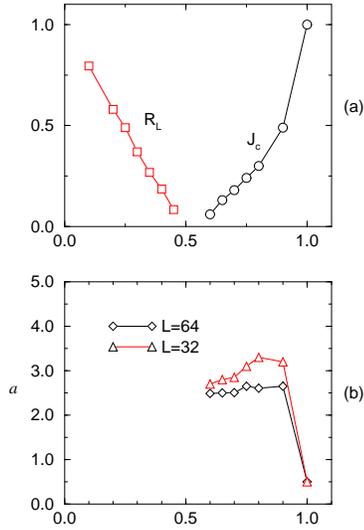,bbllx=1cm,bblly=1cm,bburx=20cm,
bbury=20cm,width=8cm}
\caption{(a) Critical current $J_c$ and linear resistance $R_L$ as a
function of $p$ for $L=64$. (b) Exponent $a$ of the power-law behavior 
$E \sim (J-J_c)^a $.}
\end{figure}

At or sufficiently close to $p_c$, where the percolation correlation length
is the dominant length scale, one expects a different behavior. In fact, by
matching the scaling of the IV curves below and above $p_c$, it has been
shown that \cite{stroud2} at $p_c$ the power-law exponent $a= 1 + s/[\nu
(d-1)] $ with $J_c=0$ which gives $a\sim 1.98$ in $d=2$. We find $a=2.1(2)$
for $L=64$ (not indicated in Figure 2b) which is consistent with this value.

Some insight into the possible universal behavior for $p>p_c$ can be
obtained by regarding the onset of resistive behavior for $J>J_c$ as a
dynamical critical phenomenon driven by the external current, where
power-law behavior appears naturally as a result of scaling. The required
scaling assumptions are similar to those proposed by Fisher {\it et al.\cite
{fisher}} in a different context. Above the critical current, the
superconducting coherence length $\xi $ is finite, leading to resistive
behavior. We assume that it diverges as a power law $\xi \sim (J-J_c)^{-{\nu
_I}}$ near $J_c$, where $\nu _I$ is a critical exponent characterizing the
current driven transition. From the definition of the electric field we have 
$E=-\partial _tA$, where $A$ is the vector potential which enters the
Hamiltonian of the JJA in the dimensionless form $\int_r^{r^{\prime }}A\cdot
dr$. The typical time scale is given by the relaxation time which diverges near
$J_c$ as $\tau \sim \xi ^z$, where $z$ is a dynamic critical exponent. 
From dimensional analysis we
then expect the scaling $E\sim 1/(\xi \tau )\sim \xi ^{-1-z}$ and a
power-law behavior of the current-voltage curve 
\begin{equation}
E\sim (J-J_c)^a,\ \qquad a=(z+1)\nu _{I\text{ \ }\ ,}  \label{ixv}
\end{equation}
above $J_{c\text{ }}$. Sufficiently close to the percolation threshold,
there is another characteristic length scale, $\xi _p\sim (p-p_c)^{-\nu _p}$%
, the percolation correlation length, and one expects a crossover to another
critical behavior \cite{stroud2}. We have only considered the case of a
single relevant length scale and focus in the regime $p>p_c$.

The exact values of the critical exponents $\nu _I$ and $z$ are not known
but the following qualitatively estimate of these exponents appear
consistent with the numerical results. We expect that at a characteristic
current density $J>J_c$, phase coherence can change significantly in a
correlation volume of order $\xi ^d$. In this volume, the typical variation
of the phase is $\sim 2\pi $. By comparing the coupling term of the external
current to the phase gradient $\int (J-J_c)\nabla \theta $ which appears in
the continuous version of the JJA Hamiltonian and the quadratic
approximation to the Josephson energy term $K^{*}\int (\nabla \theta )^2$,
where $K^{*}$ is an effective stiffness, one finds that $(J-J_c)$ should
scale as $(J-J_c)\sim 1/\xi $ and so $\nu _I=1$ . Thus, the only remaining
parameter is the dynamical exponent $z$. If we neglect non-linearities and
assume relaxation dynamics as in time-dependent Ginzburg-Landau models,
we expect $z=2$, and therefore $a=(z+1)\nu _I=3$. On the other hand,
recent work\cite{tiesinga} suggests that for the dynamics described by
Eq. (\ref{green}), where there is  current conservation at each lattice site,
the dynamical exponent is $z \sim 0.9$ and so $a=1.9$.
The data in Fig. 2b for the largest  system size $L=64$ is intermediate between
these two values. Since in Eq. (\ref{ixv}) $a$ depends both on $\nu _I$ and $z$, 
we need additional results to test this analysis.

We have performed a finite-size scaling analysis at a fixed value of $p=0.7$
to verify the scaling behavior of Eq. (\ref{ixv}) and extract an independent
numerical estimate of $\nu _I$. In a finite system, the correlation length
is limited by the system size $L$ and finite-size scaling leads to 
\begin{equation}
EL^{a/\nu _I}=f((J-J_c)L^{1/\nu _I}),  \label{ixvL}
\end{equation}
where $f$ is a scaling function. From Eq. (\ref{ixvL}), all data in the
scaling plot $EL^{a/\nu _I}\times (J-J_c)L^{1/\nu _I}$ should collapse on to
the same curve if $\nu $ and $a$ are chosen correctly as shown in Fig. 3 for
system sizes ranging from $L=16$ to $64$. We find that a reasonable scaling
behavior is obtained for $J_c\sim 0.19(2)$, $\nu_I \sim 1.1(3)$ and $a\sim
2.4(2)$ where the error estimates are obtained by averaging various
estimates of $J_c$, $\nu _I$ and $a$ which give equally acceptable scaling
plots. Using the relation in Eq. ({\ref{ixv}), we find $z=1.2(6)$. The value
of $\nu _I$ is in agreement within the errors with the one predicted above
but $z$ is not accurate enough to allow any comparison. Improved estimates
and further analytical work are necessary for a detailed study of the
critical behavior. The close analogy to other critical phenomena near
threshold \cite{nayaran} may provide an interesting approach and eventually
identify the relevant universality class.

\begin{figure}[tbp]
\centering\epsfig{file=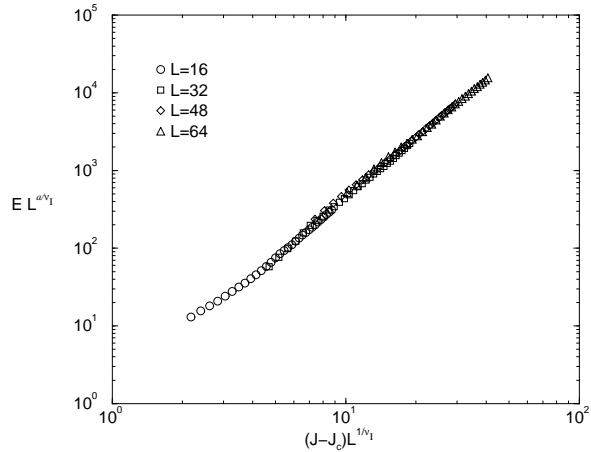,bbllx=1cm,bblly=1cm,bburx=20cm,
bbury=20cm,width=8cm}
\caption{Finite-size scaling plot of the power-law behavior $E \sim
(J-J_c)^a $ for $p=0.7$ at zero temperature, using $J_c=0.19$, $\nu_I =0.9$
and $a=2.5$.}
\end{figure}

It is interesting to note that if we assume $\nu _I=1$ in Eq. (\ref{ixv}),
the result $a=z+1$ is similar to the one inferred by Prester \cite{prester}
based on an analogy between the onset of dissipation at the critical current
and the resistance of a random resistor network which leads to $a=t+1$,
where $t$ is the conductivity exponent of a mixture of resistors and
insulators. This would suggest $z=t$. In two dimensions, where \cite
{stinchcombe} $t\sim 1.3$, this gives $z=1.3$ and $a=2.3$ which is in fact
consistent with our numerical estimates.

We now turn to finite temperature effects. We have only studied the behavior
at percolation threshold where the superconducting transition is known to
occur \cite{stinchcombe} at $T=0$. Again, this is the simplest case where
there is a single dominant length scale in the system which at finite
temperature is the thermal correlation length $\xi _T$. This correlation
length diverges for decreasing temperature as $\xi _T\propto 1/T^{\nu _T}$
where $\nu _T$ is the thermal correlation length exponent of the diluted XY
model which is isomorphic to the JJA at zero current. One can anticipate
that the increasing correlation length will have important effects in the
nonlinear resistance for decreasing temperatures \cite{fisher,eg}. At any $%
T>0$, the linear resistance, $R_L=\lim_{J\longrightarrow 0}E/J$ , is nonzero
since the superconducting correlation length is finite and is expected to be
thermally activated. However, in presence of a finite current density $J$,
an additional length scale $L_I\propto kT/J$ is set by the external current 
\cite{fisher} due to temperature fluctuations. This can be obtained by
comparing the extra energy arising from the coupling to the external
current, $JL_I\Delta \theta $ within a length scale $L_I$ and for a typical
phase variation $\Delta \theta \sim 2\pi $, to the thermal energy $kT$. For $%
\xi _T<L_I$, which holds for sufficiently small $J$ at any finite $T$, the
linear resistance is basically unchanged since the smaller length scale $\xi
_T$ dominates the activation energy. However, for current densities larger
than $J_{nl}\propto T^{1+\nu _T}$ nonlinear behavior sets in as $L_I<\xi _T$
in this case. So, the range in $J$ where $E/J$ is roughly a constant should
decrease with temperature and the characteristic current density $J_{nl}$
where it crosses over to nonlinear behavior decreases as a power law $%
T^{1+\nu _T}$ with a universal exponent. Associated with the divergent
correlation length $\xi _T$ one also defines a relaxation time $\tau $ that
owing the zero-temperature transition does not follow the usual form $\tau
\propto \xi ^z$ and can have an exponential temperature dependence. Since
the electric field scale as $E\sim 1/(\xi \tau )$, the current density as $%
J\sim kT/\xi $ and using $\xi =1/T^{\nu _T}$ the nonlinear resistance
behaves as \cite{fisher} 
\begin{equation}
\frac EJ=\frac 1{\tau \ T}\ g(\frac J{T^{1+\nu _T}})  \label{E/J}
\end{equation}
in two dimensions, where $g$ is a scaling function. If the linear resistance 
$R_L$ is finite for any $T>0$ then $g(0)$ is a constant which can be set to
unity, $g(0)=1$. When  the nonlinear resistance $E/J$ is normalized by 
its linear value $R_L$ at the same temperature one can then write 
\begin{equation}
\frac E{JR_L}=g(\frac J{T^{\nu _T+1}})  \label{ixvT}
\end{equation}
It is clear from Eq. (\ref{ixvT}) that the characteristic current at which
nonlinear behavior is expected to set in varies as $T^{1+\nu _T}$ as
mentioned before.

The nonlinear resistance $E/J$ obtained numerically at $p=p_c$ for the
largest system size $L=64$ is shown in Fig. 4. The curves shows the expected
behavior. For small current densities $J$, there is a linear contribution
where $E/J$ tends to a finite value which depends on the temperature. This
is more clear for the highest temperature $T=0.7$ where the range of $J$ in
which $E/J$ is roughly a constant is more pronounced. For increasing current
densities it crosses over to a nonlinear behavior. As temperature decreases
nonlinearity appears at smaller currents and the linear behavior is less
clear. For the lowest temperature $T=0.3$ the linear behavior presumably
occurs at current densities smaller than $J=0.02$ but numerical calculations
in this range requires very long equilibration times which prevent us to
confirm this behavior. In fact, as discussed below, the relaxation time $%
\tau $ increases very rapidly (possibly exponentially) with decreasing
temperatures.

\begin{figure}[tbp]
\centering\epsfig{file=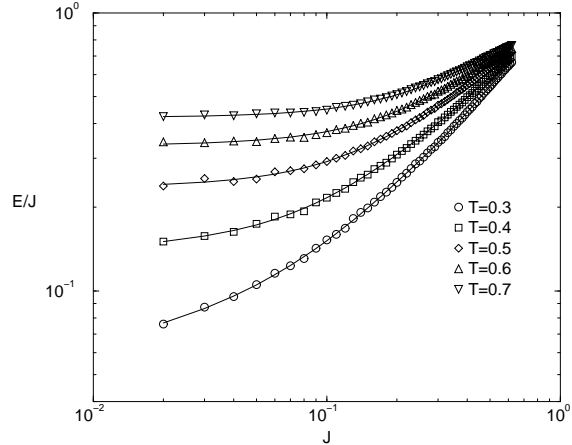,bbllx=1cm,bblly=1cm,bburx=20cm,
bbury=20cm,width=8cm}
\caption{Nonlinear resistance $E/J$ as a function of temperature for $L=64$
at $p=0.5$. Continuous lines are a guide to the eyes.}
\end{figure}

The temperature dependence of the linear resistance estimated at the lowest
current is indicated in Fig. 5a. It seems to be consistent with an activated
behavior with an energy barrier $E_b\sim 0.91(3)$. Our data is not accurate
enough and the temperature range is too limited to exclude more complicated
behavior as a temperature dependent $E_b(T)$. The links-nodes picture of the
percolation cluster \cite{stinchcombe} might suggest a single-junction
behavior which in fact gives an Arrhenius behavior at sufficiently higher
temperatures \cite{bishop} but this would give a much larger barrier of $%
E_b=2$ and requires that the Josephson coupling of the effective
single-junction is not renormalized. In addition, the scaling form in 
Eq. (\ref{ixvT}) which is found in our case as discussed below, does not 
hold for
the single junction as can be verified from the closed-form solution of the
current-voltage relation \cite{bishop}. It is unclear at the moment what is
the appropriate model to describe the temperature dependence of the $R_L$.
In any case, if the apparent exponential behavior of $R_L$ holds down to very
low temperatures it implies, from Eq. (\ref{E/J}), that the relaxation time
diverges exponentially $\tau \sim \exp (E_b/T)/T$ and so the current-voltage
characteristics at very low temperatures and currents may be inaccessible by
direct simulation.

In Figure 5b, we show the temperature dependence of $J_{nl}$ for the data in
Fig. 4 in a log-log plot. It is consistent with the power law behavior $%
J_{nl}\propto T^{1+\nu_T }$ and provides a direct estimate of the thermal
exponent $\nu _T=1.2(2)$. To estimate $J_{nl}$ we defined the crossover to
nonlinear behavior as the value of $J$ where $E/JR_L$ starts to deviate from
a fixed value  equals to $1.2$. The slope in the plot of Fig. 5b should not
dependent on this value as long as it is not too large compared to unity and we
checked that other choices gives the same results within the error
estimates. In Figure 6, we show the scaling plot of the data in Fig. 4
according to the scaling behavior of Eq. (\ref{ixvT}). The scaling plot is
obtained by adjusting a single parameter $\nu _T$ that gives the best data
collapse. This is consistent with the scaling behavior discussed above and 
gives an independent estimate of $\nu _T\sim 1.4(2)$. From these results we 
obtain a final estimate of the thermal critical exponent $\nu _T=1.3(3)$. 
This value of $\nu _T$ is in fact consistent with the thermal correlation 
exponent of the diluted XY model at percolation threshold \cite{stinchcombe}, 
$\nu=0.98-1.03$.

\begin{figure}[tbp]
\centering\epsfig{file=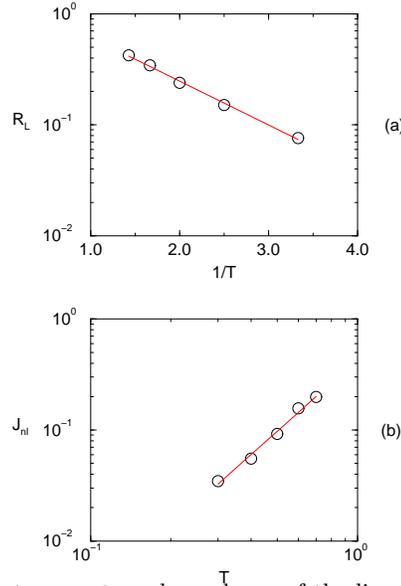,bbllx=1cm,bblly=1cm,bburx=20cm,
bbury=20cm,width=8cm}
\caption{ (a) Arrhenius plot for the temperature dependence of the linear
resistance $R_L$, estimated at $J=0.02$, for $L=64$ at $p=0.5$. (b)
Crossover current density $J_{nl}$ where nonlinear behavior appears as a
function of temperature. The slope gives and estimate of $1+\nu_T=2.2(2)$}
\end{figure}

Our analysis for the temperature effects  on the nonlinear resistance is 
strictly valid at $p_c$, where the percolation correlation length $\xi_p$
is infinite but the thermal correlation length $\xi_T$ is finite. In order
to compare to the available experimental data of Harris {\it et al.} 
\cite{garland} on artificial arrays for $p$ close to percolation threshold, 
which is in fact consistent with a vanishing transition temperature,
we must  take into account the competing effects of $\xi_T$ and $\xi_p$ which 
is a more complicated problem.  However, sufficiently close to $p_c$ the
analysis should still be valid at high temperatures when $\xi_T << \xi_p$. 
Unfortunately, the scattering of the data at small currents and the limited
range of temperatures  where linear resistive behavior is apparent  prevent 
us to perform the same scaling analysis as described above. 

\begin{figure}[tbp]
\centering\epsfig{file=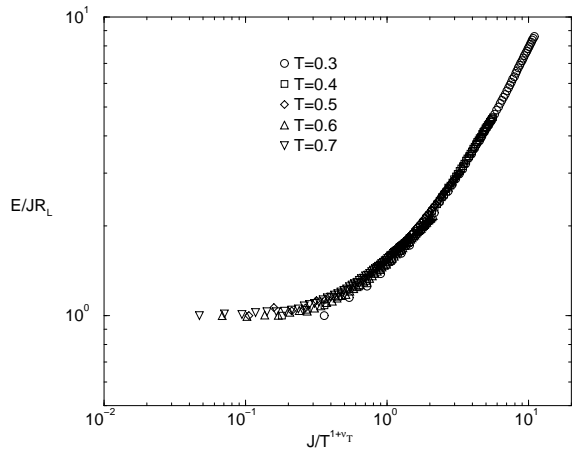,bbllx=1cm,bblly=1cm,bburx=20cm,
bbury=20cm,width=8cm}
\caption{ Scaling plot of the data in Fig. 4 for $\nu_T = 1.4$.}
\end{figure}

\section{Conclusion}

In summary, we have studied the IV characteristics in a model of resistively
shunted two-dimensional diluted JJA, at zero external field, by numerical
simulations and scaling arguments. At $T=0$, the IV curves show power-law
behavior above a critical current density which decreases with dilution. The
power-law behavior follows from a simple scaling analysis which leads to $%
a=(z+1)\nu _I$, where $z$ is the dynamical exponent and $\nu _I$ is the
superconducting correlation length exponent. Numerically we find $\nu
_I=1.1(3)$ , consistent with a scaling argument which gives $\nu _I$ $=1$ ,
and $a=2.4(2)$ . The value of $a$ is in agreement with the relation $a=t+1$
, in two dimensions, which has been proposed \cite{prester} in relation to
granular high-$T_c$ materials in zero field. At the percolation threshold
and finite $T$, the results are consistent with the scaling behavior of a $%
T=0$ superconducting transition. Crossover to non-linear behavior appears at
currents proportional to $T^{1+\nu _T}$, where $\nu _T$ is a correlation
length exponent for the diluted XY model at percolation threshold. The
behavior at percolation threshold is analogous to the zero-temperature
vortex glass of disordered superconductors in a magnetic field, except for
the value of $\nu _T\ $. For experiments in arrays and granular high-$T_c$\
materials, this behavior clearly demonstrates the importance of carefully
comparing the expected power-law behavior of IV characteristics resulting
from field-induced effects to the zero-field case.

\acknowledgments

We thank S. Shenoy and M. Prester for helpful discussions. The authors
acknowledge the support from International Centre for Theoretical Physics,
where part of the work was done.

\end{document}